\documentclass[prd,nobibnotes,nofootinbib,showpacs,12pt]{revtex4}

\usepackage{amsmath}    
\usepackage{graphicx}   
\usepackage{verbatim}   
\usepackage{color}      
\usepackage{subfigure}  
\usepackage{hyperref}   

\begin{document}

\title{Remarks on correlators of Polyakov Loops}
\author{Herbert Neuberger\footnote{Weston Visiting Scientist at 
The Weizmann Institute of Science.}}
\email{neuberg@physics.rutgers.edu}
\affiliation{Department of Physics and Astronomy, Rutgers University,\\ 
Piscataway, NJ 08855, U.S.A} 

\begin{abstract}
Polyakov loop eigenvalues and their $N$-dependence
are studied in 2 and 4 dimensional SU(N) YM theory. The connected 
correlation function of the single eigenvalue distributions of two separated Polyakov loops
in 2D YM is calculated and is found to have a structure differing 
from the one of corresponding hermitian random matrix ensembles. 
No large $N$ non-analyticities are found for two point functions
in the confining regime. Suggestions are made for situations in which large-$N$ 
phase transitions involving Polyakov loops might occur. 
\end{abstract}

\date{May 2, 2013}


\maketitle

\section{Introduction}

This work is concerned with $SU(N)$ YM theory (QCDN) in 4D. 
QCDN admits a large $N$ expansion~\cite{thooft}. 
Lattice work 
has shown that there is confinement at finite and infinite $N$~\cite{lucini_rev}. Then, 
QCDN at $N=\infty$ (QCD$\infty$) 
is similar to the starting point of the topological expansion (TE)
~\cite{te}.
In TE one constructs iteratively an S-matrix from a set of postulated basic general
properties. Another starting point for the TE is provided by string theory. 
In both cases one starts from a system consisting of an infinite set of stable 
particles interacting weakly at linear order. Upon iteration,
other singularities build up. The expansion is organized 
diagrammatically with an order given by the genus
of a Riemann surface.

The QCDN route is better founded than the string one. 
We can safely assume that there exist Wightman $n$-point functions
of local gauge invariant observables that admit a single
valued continuation to the extended tube ${\cal T}'_{n-1}$~\cite{wightman} 
for any $N$. These functions determine the leading nontrivial term in $\frac{1}{N}$ 
of any amplitude entering the S-matrix. From this off shell starting point 
one might be able to build a better founded QCDN  
string theory~\cite{veneziano}. Concretely, one would need explicit forms 
of a least some of the sets of entries of the S-matrix.

Despite quite a few papers which achieved high levels of popularity, there is 
not one non-perturbative 
physical number that has been analytically calculated, or, at least credibly estimated, 
in QCDN (with or without a finite number of quarks) at leading order in $\frac{1}{N}$ 
or $\frac{1}{N^2}$. Nevertheless, interest in large $N$ does
not seem to die out. Quite a few workers, me included, 
still are trying to get some new quantitative result in QCDN 
which rests on the simplification afforded by $N\gg 1$. 

My idea has been to find a simple physical  
single scale observable whose behavior as a function of this scale showed
a universal behavior at the crossover separating long from short scales.
Large $N$ comes in to provide this universality by a large $N$ phase transition. 
The universality then becomes
a random-matrix type of universality. The hope is to exploit it
in order to match an effective string descriptions holding at large distances
to perturbation theory holding at short distances. 
For example, consider a circular Wilson loop of radius
$r$. For $r$ large effective string theory provides 
some universal information about the $r$ dependence, while at small $r$ 
perturbation theory applies; the new ingredient is 
that random matrix universality would provide the
means to connect these two dependencies. The hope is that 
an approximate connection between the string tension and some standard
perturbative scale would then be calculable. The existence of the large 
$N$ phase transition is believable for the circular loop because
it has been established numerically for square loops. However, 
it would be preferable to consider smooth loops also on the lattice, and 
this leaves us with only Polyakov loops winding around a compactified 
direction. The length of this circle has to be bounded from below in order to
stay in the confined phase. The single eigenvalue density, $\rho^{(1)}$, of a Polyakov
loop becomes uniform at $N=\infty$ on account of the well known $Z(N)$ symmetry.
This leaves us with $\rho^{(2)}$, the connected correlation function of the $\rho^{(1)}$'s
of two separated Polyakov loops, as the simplest smooth observable on the lattice. 

In this paper, I focus on Polyakov loops. The outline of the papers is:
Sections \ref{vol_red} and \ref{two_loop_a} provide background material.
The concrete new results are in \ref{two_loop_b}. They consist of
an evaluation of the single Polyakov eigenvalue density connected 
two point correlation function under the assumption of second rank Casimir dominance.
A formula for any $N$ (taken as odd, for simplicity) is provided, the large $N$ 
limit is taken and the validity of the latter is checked numerically.
Next, a brief comparison with Monte Carlo data in four dimensional $SU(N)$ Yang Mills
theory is carried out. There are no large $N$ phase transitions. Incidentally
it is noted that the result
does not show universal features known to hold for large hermitian matrix ensembles.
Section \ref{other} contains ideas for future work. A short summary concludes the
paper. 

\section{Volume Reduction\label{vol_red}}

QCDN is a field theory, but geometrically the fundamental variables are not fields 
defined over ${\cal M}=R^4$, but rather fields defined over loops in $R^4$. This becomes
particularly evident when one introduced a lattice UV cutoff: One can derive an
infinite set of equations connecting various loop operators and the equations 
reflect the ordinary locality of four-space the field theoretical 
formulation rests on, without any of the collateral expenditures (gauge fixing, Faddeev-Popov ghosts, 
Gribov ambiguities) associated with
formulating the continuum 
theory in terms of gauge fields~\cite{makeenko_migdal}. 
The loop equations self-truncate at infinite $N$, feeding the
hope that it ought to be easier to handle non-perturbative issues of QCDN at $N=\infty$
~\cite{makeenko_migdal}.
Taking the equations to the continuum is hampered by the nonexistence of anything
resembling a decent calculus in loop space. One way to go around this obstacle is to try to guess a 
well defined solution 
directly in the continuum (which obeys 
general symmetry/unitarity constraints) 
and show that it satisfies a set of equations that can be viewed as 
a concrete realization of the formal continuum 
loop equations~\cite{polyakov}. This has led to progress
in string theory and even to a connection back to field theory, 
but not for QCDN~\cite{maldacena}. 
As far as I know, we still do not have even one nontrivial example where the formal 
loop equations have been credibly defined in the continuum. 

One consequence of the loop equation is that at $N=\infty$ the replacement of $R^4$ 
by $T^4$, where the sides of the torus are all larger than the
inverse deconfinement temperature, preserves a large subset of observables with no
dependence on the actual finite length of these 
sides~\cite{cont_reduction}. This is of some
help in numerical work, but the saving is 
quite limited~\cite{rect_w_loops}. The term describing
this phenomenon is ``reduction'', on account of a reduction in the number 
of degrees of freedom as far as the four volume size goes. Reduction can
be applied to any number of directions and one assumes that there is 
a hierarchy of scales associated with the preservation of the associated
$Z^k(N)$, $k=1,2,3,4$~\cite{cont_reduction}.   

The restriction on the sides of $T^4$ ensures that the global $Z^4(N)$ symmetry
now present is not broken spontaneously by the $N=\infty$ limit. The preservation
of the consequence of this symmetry on expectation values of parallel 
transport round non-contractible loops is a necessary ingredient 
for reduction~\cite{ek}.
The equivalence between the $R^4$ and $T^4$ loops equations breaks down if
$Z^4(N)$ is not obeyed by expectation values of winding loops at $N=\infty$
~\cite{qek}. 
If the $Z^4(N)$ is preserved at $N=\infty$, the spacings 
between momenta induced by the
finite volume of $T^4$ get continuously filled in by the 
eigenvalue-sets of winding loops~\cite{cont_reduction}. 

Polyakov loops are the natural extra observables one has when considering 
${\cal M}=S\times R^3$. We assume that the compact direction is large enough that
the one $Z(N)$ is preserved at all $N$. When we slice ${\cal M}$ by 
three-spaces parallel, or orthogonal, to the compact direction 
we find two different transfer matrices. They provide two Hamiltonian
pictures. In one picture 
$Z(N)$ is just an extra global symmetry of the Hilbert space and the
system is at zero temperature. In the other, the system is at finite temperature.
In either case, one has a Hilbert space 
which can be chosen to transform irreducibly under 
the symmetries commuting with the Hamiltonian. 
The spaces and Hamiltonians are different, providing different spectral representations of
identical observables.

Reduction applies only to non-winding loops; it is of interest to see if any 
reduction-related simplifications hold at $N=\infty$ also for winding loops
~\cite{neuberger_ham}.

\section{Correlations of two Polyakov loops\label{two_poly}}

Define the parallel transport winding loop operator by:
\begin{equation}
U_P (x)={\cal P} e^{i\oint_{x_4}^{x_4} A_4 ({\vec x},\tau) d\tau}
\end{equation}
The compact direction is 4. $A_\mu(x)$ are the gauge fields given by traceless
hermitian $N\times N$ matrices and $({\vec x})_i=x_i, i=1,2,3$. ${\cal P}$ is path
ordering. Polyakov loops are independent of $x_4$:
\begin{equation}
P_R ({\vec x})= \frac{1}{d_R} \chi_R (U_P(x))
\end{equation}
$R$ labels an irreducible representation of $SU(N)$, $\chi_R$ is the character in $R$
and $d_R$ is the dimension of $R$. If the number of boxes in the Young pattern corresponding
to $R$ is $m_R$, the $N$-ality of $R$ is given by $n=\mod(m_R,N)$. Under $Z(N)$ we have
$P_R \to e^{2i\pi n /N} P_R$. 

Consider the case $n\ne 0$. Then $\langle P_R\rangle=0$ and $G_R(r)=\langle P_R(0) P_{\overline R} (r)\rangle$
is generically nonzero. Here ${\overline R}$ is conjugate to $R$. $r$ is a positive distance. Denoting 
the length of the compact direction by $l$, the two point function is a function of $R$ and the two length
scales $l,r$. Define (formally) $W_R(l,r)=\log G_R(r)$. We assume that the $\theta$-parameter 
in front of the $\int d^4 x Tr[F\wedge F]$ parameter in the action is set to zero,
so $G_R(r)$ is real and positive. The definition is only formal because after renormalization
there will be an arbitrary term in $W_R(l,r)$ of the form $\mu_R l$. 
Thus, $\partial W(l,r)/\partial l$ is well defined up to an additive constant
~\cite{polyakov_w_loop}. 

\subsection{The simplest asymptotic properties\label{two_loop_a}}

Since $l > 1/T_c$ where $T_c$ is the deconfinement temperature, the first reaction
would be that $W$ cannot be computed in perturbation theory. Nevertheless, 
to some extent, the quantity 
\begin{equation}
\lim_{l\to\infty} \partial^2 W(l,r)/\partial l \partial r= dV_R(r)/dr
\label{pot}
\end{equation}
can. Instead of the $l\to \infty$ limit in eq. (\ref{pot}) one takes an infinite 
uncompactified $x_4$ axis and replaces $l$ by $T$ and $W(l,r)$ by
the logarithm of a rectangular Wilson loop of shape $T\times r$. 
Inspection 
of Feynman diagrams shows that in the $T\to \infty$ 
limit one has an expansion for $V_R(r)$ up to two loop
order in the coupling which
ought to be useful for $r$ small enough. 
To go to higher loops one needs to include non-analytic terms in
the coupling~\cite{appel_dine_muz}. pNRQCD provides a prescription 
for how to do this, but I doubt that it is unique~\cite{pineda}. 

In this expansion the path integral 
over $A_4$ is expanded around $A_4=0$, which breaks
the $Z(N)$. This may not matter for the two point Polyakov loop 
function at infinite $l$. Whether it can be used at any finite $l$ 
is an open question, so long as $l > 1/T_c$. 
If there were a credible perturbative regime in which the $l$-scale can
somehow be removed from the problem, one would expect that $\rho^{(2)}$
might show some crossover as $r$ is varied. Then, the ingredients necessary
for a large $N$ phase transition to develop are present. The hope is that, 
for $r$ small enough, the eigenvalues of one Polyakov loop would restrict
the fluctuations of the eigenvalues of the other Polyakov loop to such a degree that the periodicity in the
angle differences would barely be felt. For $r$ large this periodicity would
get restored to full strength. A separating crossover at finite $N$ would
become a phase transition at infinite $N$. The intuition behind the focus 
on eigenvalues is that collectively their fluctuation 
explore the distance of parallel transport operators from unity. Only
when the compactness of the group is felt does one expect non-perturbative
effects to become important. Compactness is felt only when parallel 
transport exceeds a certain distance from unity.
We shall see that the hope for a transition is not realized. 

Beyond perturbation theory, $V_R(r)$ is the ground 
state energy of the Hamiltonian associated
with evolution in the $x_4$ direction in the sector 
defined to transform under the local 
gauge group as $R$ at ${\vec r}=(0,0,0)$ and ${\overline R}$ 
at ${\vec r}=(r,0,0), r>0$. This 
ground state is $d_R^2$ degenerate. When one computes the 
partition function viewing
$l$ as the inverse temperature, the degeneracy of 
the ground state cancels the 
prefactors normalizing the Polyakov loop operators. 
There is no overall factor of $N^2$ 
in the physical piece of the free energy. 

Because of the representation content of the Hilbert space which
breaks translation invariance, there is no physical
interpretation in this picture for plane waves propagating in the 
$x_{1,2,3}$ directions superposing ground state states. 
\footnote{In pNRQCD one deals with an expansion in inverse 
quark mass; in this context it is possible to provide
a meaningful definition of the spatial Fourier Transform of $V(r)$
because the sources can move. At three loop order the expansion in powers of the strong-force 
coupling breaks down, but the non-analytic term responsible for this (in this
framework) can
be derived. } 
The dependence of $dV(r)/dr$ on 
$r$ for $r\to\infty$ starts with a constant (the string 
tension of open strings with fixed endpoints
transforming as $R$ and ${\overline R}$ respectively) and 
continues to subleading orders; several terms
in this expansion are universal. I shall describe below in more 
detail this aspect in another asymptotic limit.

One can also consider the large $r$-separation fixed $l$ limit:
\begin{equation}
{\cal F}_R(l)=\lim_{r\to\infty} \partial^2 W_R(l,r)/\partial l \partial r
\end{equation}
Confinement in this context means that 
$\lim_{l\to\infty} {\cal F}_R(l) =\sigma_n > 0$, with
$\sigma_n$ depending only on the assumed non-zero $N$-ality 
of $R$, $n$; $\sigma_n=\sigma_{N-n}$. 
${\cal F}_R(l)$ gives the $l$-derivative of the $r$-derivative at $r=\infty$ of eigenvalues
of the Hamiltonian describing
evolution in any one of the directions $x_j, j=1,2,3$ in the 
$n$-sector of the global $Z(N)$. Now, it does make sense
to superpose states and project on zero spatial momentum. This
gives the ground state energy in the $n$-winding sector.
One can look at several subleading terms in the large $l$ 
expansion. For this define 
${\cal F}_R(l) = \sigma_n {\hat F}_R(l\sqrt{\sigma_n})$
\begin{equation}
{\hat F}(x)=1+c_1/x^2+c_2/x^4+c_3/x^6+...
\end{equation}
Assuming an effective string theory description one derives 
from symmetry principles alone that
the coefficients $c_{1,2,3}$ are universal calculable 
finite numbers, independent of $R$ (and consequentially
of $n$). They are actually also independent of any other 
detail regarding the field theory, except
the assumption of confinement and applicability of 
effective string theory~\cite{ofer_zohar}. 

To summarize: at any finite $N$, $W_R(\infty,r)$ and $W_R(l,\infty)$ with $R$ of
nonzero $N$-ality have some universal coefficients 
in their asymptotic expansions in $r$ and $l$ respectively, which follow from the assumption of
confinement and applicability of effective string theory. Numerical checks have yielded
results consistent with this. The $N\to\infty$ limit provides no further simplification
with respect to these properties.~\footnote{One should distinguish between 
the dream string theory which is equivalent to QCDN (it is unknown whether such a dream string 
theory actually exists) and the effective string theory
we are talking about; the dream (closed-) string theory would have a coupling constant which 
goes as $\frac{1}{N^2}$ and the $N\to\infty$ limit would turn 
the interactions off. One would need the $N\to\infty$ limit in order to justify
the focus on the lowest genus surface relevant to the correlation under investigation.
The effective string theory on the other hand has already summed up all 
contributions from handles of the dream string theory, and the universal
results provided by the cylinder are $N$-independent.}
There is a hope that the $N\to\infty$ limit could
provide a clear demarcation point for the domain in which the asymptotic expansion
of long strings can make any sense at all, 
but the universal predictions of effective string 
theory are insensitive to this. 

\subsection{Looking for large $N$ phase transitions\label{two_loop_b}}

Effective string theory can also be applied to the case where $l$ and $r$ are both 
taken to infinity; a brief overview with references to 
original work can be found in~\cite{rect_w_loops}. This can be done 
also for ordinary contractible rectangular Wilson loops 
in $R^4$. Reduction applies in these cases. There is no lower limit on $l$ or $r$ and
therefore there is a perturbative regime. Renormalization to remove perimeter divergences
is still required. Again, for individual irreducible representations large $N$ provides
no additional constraints on the universal large $l,r$ results of effective string theory.
However, now there is an evident connection to a 
perturbative regime where both $l$ and $r$
are small. This connection is smooth. Only by looking at many 
representations simultaneously
does one detect a large $N$ transition: one finds a non-analyticity 
in the single eigenvalue
distribution at a point which serves as a boundary between the perturbative and
non-perturbative regime~\cite{ev_fourd}. 
The boundary location depends on an arbitrary dimensional smearing parameter. 
This parameter is a remnant of the 
need to eliminate the perimeter
divergence. It is defined in a manner independent of $R$~\cite{smearing}. 
Only at infinite $N$ is there a well defined transition point. For any finite $N$
one has only a cross-over. Numerical and analytical considerations indicate that the
large $N$ transition has a certain random-matrix model universality~\cite{ev_fourd}. 

At any finite $N$ the Polyakov loop two point function has a discontinuity as $l$ is varied through
$l=\frac{1}{T_c}$ separating the confined and deconfined phases. Since one 
can compute $V_R(r)$ in perturbation theory
at $l=\infty$, one may hope that a specifically
\underline{large $N$} transition as $r$ is varied takes place at any $l$ in the confined phase.

Lattice checks of predictions of effective string theory have overall 
been successful, at times even surprisingly successful; this
has nothing to do with $N$ being large enough. Just like effective Lagrangians
for massless pions work already on the lattice since these Lagrangians can 
parametrize any reasonable UV behavior, effective string theory should also
hold directly on the lattice so long as one is in the rough phase. 
For long strings the lattice violation of $O(4)$ invariance makes 
an impact determined by symmetries. If we replace the hypercubic
lattice by an $F_4$ lattice~\cite{ffour}, the effective 
string theory predictions may apply even better. So, the success of lattice checks of effective 
string theory might be ``explained'' by the relative unimportance of the
proximity of the field theoretical continuum limit. For example, QCDN with
scalar matter fields in the adjoint has no continuum limit; on the lattice
it would confine and effective string theory would make some universal 
predictions in the rough phase of the loop under consideration.
 
In general, strings are
less sensitive to high mode cutoffs than field theories are~\cite{marek}.
In four dimensions there is an exception~\cite{marek} which 
has to do with the fluctuations in the extrinsic string curvature. 
Potentially significant deviations from effective string theory were 
observed in~\cite{rect_w_loops} and independently in~\cite{arroyo};
in~\cite{rect_w_loops} these deviations were tentatively attributed 
to the corners of lattice loops because of the perturbation theory
experience with the corner divergences in Wilson loops.

The technical reason for the universality of 
the large $N$ transition in the case of contractible Wilson loops 
has been second rank Casimir dominance of the dependence\footnote{
The expectation value of contractible Wilson loops of all sizes 
can be written as exponents of expressions
in which the dependence on the representation enters predominantly 
through the second rank Casimir. In order to get eigenvalue distributions
one needs to sum over a set of representations. The factors 
multiplying the Casimirs are smooth in the geometrical parameters of the loop.
This structure can induce large $N$ phase transitions.} on the representation $R$. 
One may then map the
size parameters that are varied into an appropriate measure of separation in 2D YM.
So, we ask whether there is a large $N$ phase transition to be found in an analysis
of Polyakov loop correlators in YM on a 2D cylinder. 

\subsubsection{Eigenvalue-eigenvalue correlation function in 2D YM}

In general, it is known that there are large $N$ non-analyticities 
in 2D YM on a cylinder~\cite{gross_matytsin}. 
Here we wish to see if they show up in the two point function of the single eigenvalue
distributions associated with two separated Polyakov loops. In the context
of large size hermitian matrices two point single eigenvalue correlation functions have
been shown to have some universal properties~\cite{eynard}. 
A side result of the calculation below
will be to check whether this universality extends to the simplest 
unitary matrices' ensemble
with a global $Z(N)$ symmetry. From experience with hermitian matrix models we expect
smooth and strongly fluctuating contributions to the two point eigenvalue function
of equal magnitude; the oscillating piece has to be first separated out and only the
remaining smooth piece can exhibit a universal large $N$ phase transition. 

In the subsequent 
equations the assumption that $N$ is odd is made implicitly. The partition function 
of $SU(N)$ YM on a 2D
cylinder connecting two loops, $1$ and $2$, is given by~\cite{gross_matytsin}:
\begin{equation}
Z_N(U_{P_1},U_{P_2}|t)=\sum_R \chi_R (U_{P_1}) e^{-\frac{t}{2 N} C_2(R)} 
\chi_{\overline R} (U_{P_2})
\end{equation}
$t$ is the area in some area unit. Let $\rho^{(1)}_1(\alpha)$ and $\rho^{(1)}_2(\beta)$ be the single
eigenvalue distributions associated with the $N\times N$ unitary matrices $U_{P_{1,2}}$ 
respectively~\cite{single_ev_dist}:
\begin{equation} 
\rho^{(1)}(\theta; U)=\frac{2\pi}{N} \sum_{k=1}^N \delta_{2\pi} (\theta-\theta_k)
\end{equation}
The $\theta_k$ are the eigenvalues of $U$ and $\delta_{2\pi}$ 
is the $2\pi$-periodic delta-function.
The character expansion of $\rho$ is:
\begin{equation}
\rho^{(1)}(\theta;U)=1+\frac{1}{2N} \lim_{\epsilon\to 0^+} \sum_{p=0}^{N-1}\sum_{q=0}^\infty
(-1)^p e^{-\epsilon(p+q+1)} [ e^{i(p+q+1)\theta} \chi_{(p,q)} (U) +
e^{-i(p+q+1)\theta} \chi_{\overline{(p,q)}} (U)]
\label{rhoone}
\end{equation}
The irreducible representation $(p,q)$ has a Young 
pattern in the shape of a width-one hook, with
$1+p$ rows and $1+q$ columns. We wish to calculate the 
connected two point function
\begin{equation}
\langle \rho^{(1)}_1(\alpha)\rho^{(1)}_2(\beta)\rangle_c=\int 
dU_{p_1} dU_{p_2} \rho^{(1)}_1(\alpha)\rho^{(1)}_2(\beta) 
[Z_N(U_{p_1},U_{p_2}| t) -1]
\end{equation}
$dU$ is the Haar measure on $SU(N)$. In order for a 
pair $(p,q)_1$ and $(p,q)_2$ to contribute
we need that there be a singlet in their direct product. 
As $N$ is odd, this will happen only
when one pair is the conjugate of the other. For odd $N$ 
there are no $(p,q)$ self-conjugate
pairs. For $t >0$ we have:
\begin{equation}
\langle \rho^{(1)}_1(\alpha)\rho^{(1)}_2(\beta)\rangle_c =\frac{1}{N^2} 
\sum_{p=0}^{N-1} \sum_{q=0}^\infty
(-1)^p e^{-\frac{t}{2N} C(p,q) }\cos[(p+q+1)(\alpha-\beta)]
\end{equation}
Here~\cite{single_ev_dist},
\begin{equation}
C(p,q)=(p+q+1)(N-\frac{p+q+1}{N} +q-p)
\end{equation}
One can reorganize the sums to get
\begin{equation}
N^2\langle \rho^{(1)}_1(\alpha)\rho^{(1)}_2(\beta)\rangle_c \equiv 
\Re{\cal J}={\cal J}_a + {\cal J}_b
\end{equation}
The real part is taken using eq. (\ref{rhoone}) with fixed
small $\epsilon$ and subsequently setting $\epsilon=0$. 
\begin{align}
&{\cal J}_a =\sum_{n=1}^N e^{-\frac{t(N-1)}{2N^2} n (n+N)} 
\frac{1-(-1)^n e^{\frac{t}{N} n^2}}
{1+ e^{\frac{t}{n}}}\cos(n(\alpha-\beta))\\
&{\cal J}_b =\sum_{n=N+1}^\infty e^{-\frac{t(N-1)}{2N^2} n (n+N)} 
\frac{1-(-1)^N e^{tn}}
{1+ e^{\frac{t}{n}}}\cos(n(\alpha-\beta))
\end{align}
${\cal J}_a$ contains the two-point connected 
correlations $\langle Tr(U^k_{p_1}) Tr(U^{\dagger k}_{p_2}\rangle(t)$
for $k=1,..,N$. 

To obtain the large $N$ limit we write an integral 
representation~\cite{single_ev_dist}, exploiting the quadratic nature of
the $C(p,q)$ dependence on $p,q$:
\begin{align}
{\cal J}=\frac{Nu}{t} e^{-\frac{t}{2}(1-\frac{1}{N^2})} \int \frac{dxdy}{2\pi} 
e^{-\frac{N}{2t}(x^2+y^2)+\frac{(x+iy)^2}{2t}}\frac{1+u^N e^{-N(x+\frac{t}{2})+\frac{t}{2}+\frac{t}{N}}}
{1+u e^{-(x+\frac{t}{2})+\frac{t}{2N}+\frac{t}{N^2}}}\frac{1}{1-ue^{iy-\frac{2}{2}+\frac{t}{2N}+\frac{t}{2N^2}}}
\end{align}
Here, $u=\exp[i(\alpha-\beta)]$. The $y$ integral can be done by saddle point with $y_{sp}=0$.  To leading
order in $N$ we get
\begin{align}
{\cal J}\approx \sqrt{\frac{N}{t}} ue^{-\frac{t}{2}+\frac{t}{2N^2}} \int\frac{dx}{\sqrt{2\pi}} 
e^{-\frac{N}{2t} x^2 +\frac{1}{2t} x^2} \frac{1+u^N e^{-N(x+\frac{t}{2})+\frac{t}{2}}}{1+ue^{-x-\frac{t}{2}+
\frac{t}{2N}}}\frac{1}{1-ue^{-\frac{t}{2}}}
\end{align}
We need to keep the factor $e^{t/2N}$ in the denominator of the first fraction in the integrand
in order to ensure its regularity for $x+t/2 <0$, because then the numerator divides evenly by the denominator.
Carrying out the integral by saddle point we find two contributions, corresponding to the saddle
points $x^{1}_{sp}=0$ and $x^{2}_{sp}=-t$. The large $N$ limit is taken at finite non-zero $\Im (u)$ 
and finite 
positive $t$. The limits $\Im (u)\to 0$ and $t\to 0$ do not commute with the limit $N\to\infty$. 
The final results for the leading large $N$ behavior comes out to be:
\begin{align}
\Re{\cal J} \approx \frac{1}{2}\frac{\sinh\frac{t}{2} \cos\phi + 
e^{\frac{t}{2}}\left ( \sinh\frac{t}{2}\cos N\phi - \sin\phi \sin N\phi\right )}{\sinh^2\frac{t}{2}+\sin^2\phi}
\end{align}
Here, $\phi=\alpha-\beta$. The result is the sum of a smooth term and a rapidly
oscillating one. There are no large $N$ phase transitions. 

The non-oscillating term is:
\begin{equation}
\Re{\cal J}_{\rm non-oscillating}\approx \frac{1}{2}\frac{\sinh\frac{t}{2} 
\cos\phi }{\sinh^2\frac{t}{2}+\sin^2\phi}
\end{equation}
It does not exhibit the universal structure seen in continuous chains of large
hermitian matrix models~\cite{eynard}. 
Most likely, the main difference is that 
for the unitary matrix ensemble we solved above there is no 
analogue of the nontrivial potential $\int dt Tr V(M(t))$ term in the action 
for the hermitian $M(t)$ matrices. While the explicit form of $V$ is irrelevant, 
its mere presence is. 

To get a feel for the goodness of the large $N$ limit and also check whether its derivation
was correct, we present some figures below. We see that the large $N$ approximation deteriorates when $N$ decreases,
when $\phi$ is close to $k\pi, k\in Z$ and when $t$ is small, but otherwise
holds well.

\begin{figure}[h!]
\begin{center}
\subfigure[N=11, t=0.3]{\includegraphics[scale=0.4]{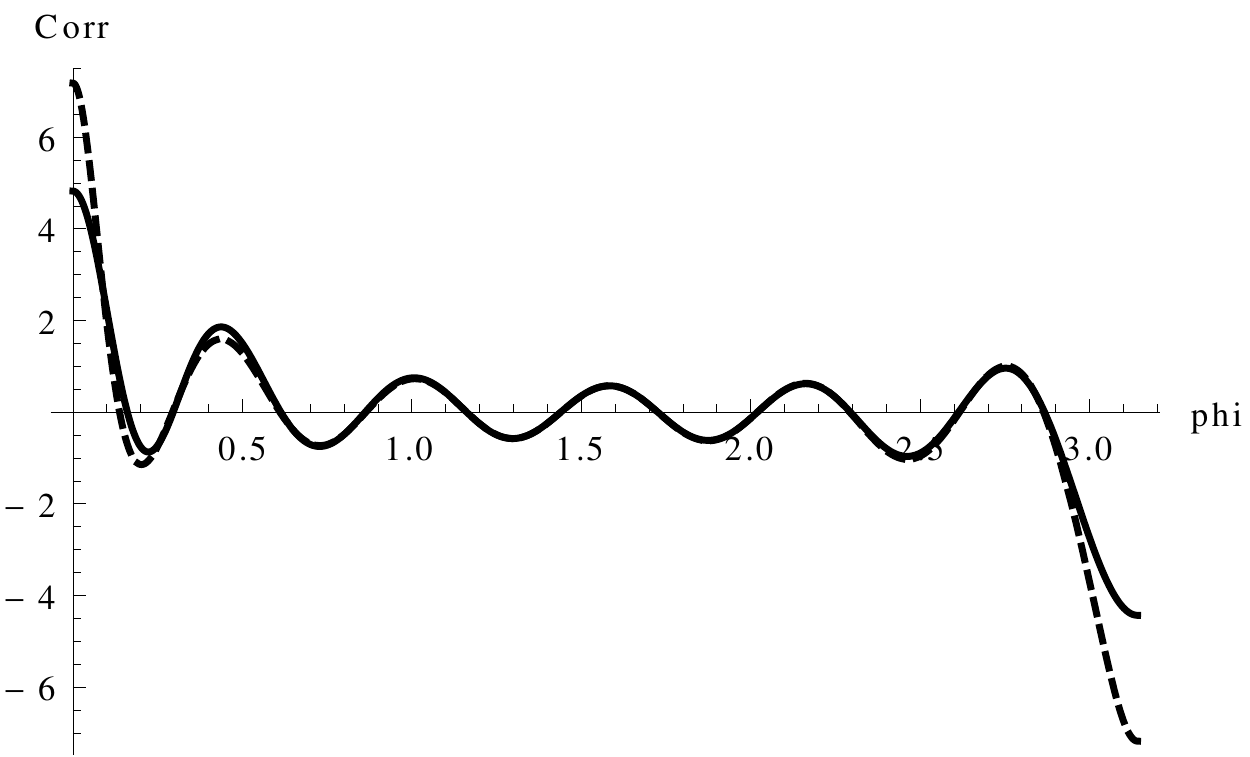}}
\subfigure[N=11, t=1]{\includegraphics[scale=0.4]{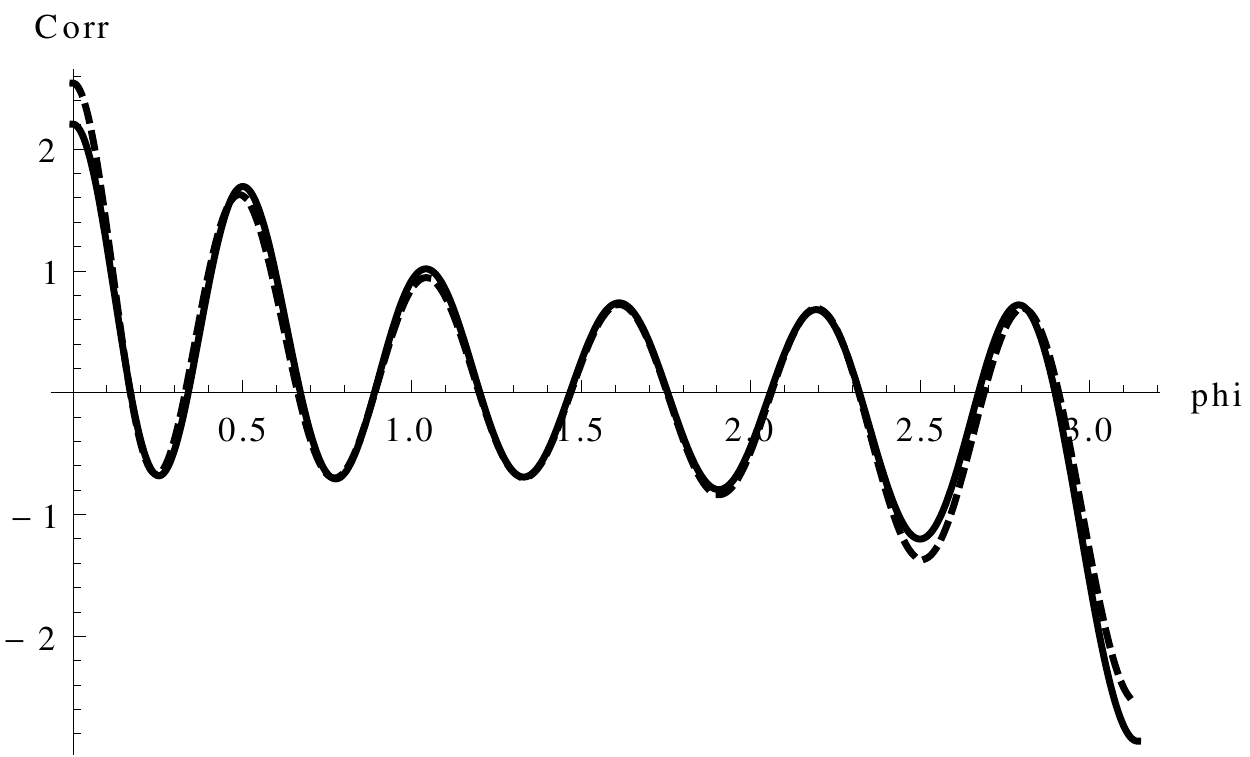}}
\subfigure[N=11, t=5]{\includegraphics[scale=0.4]{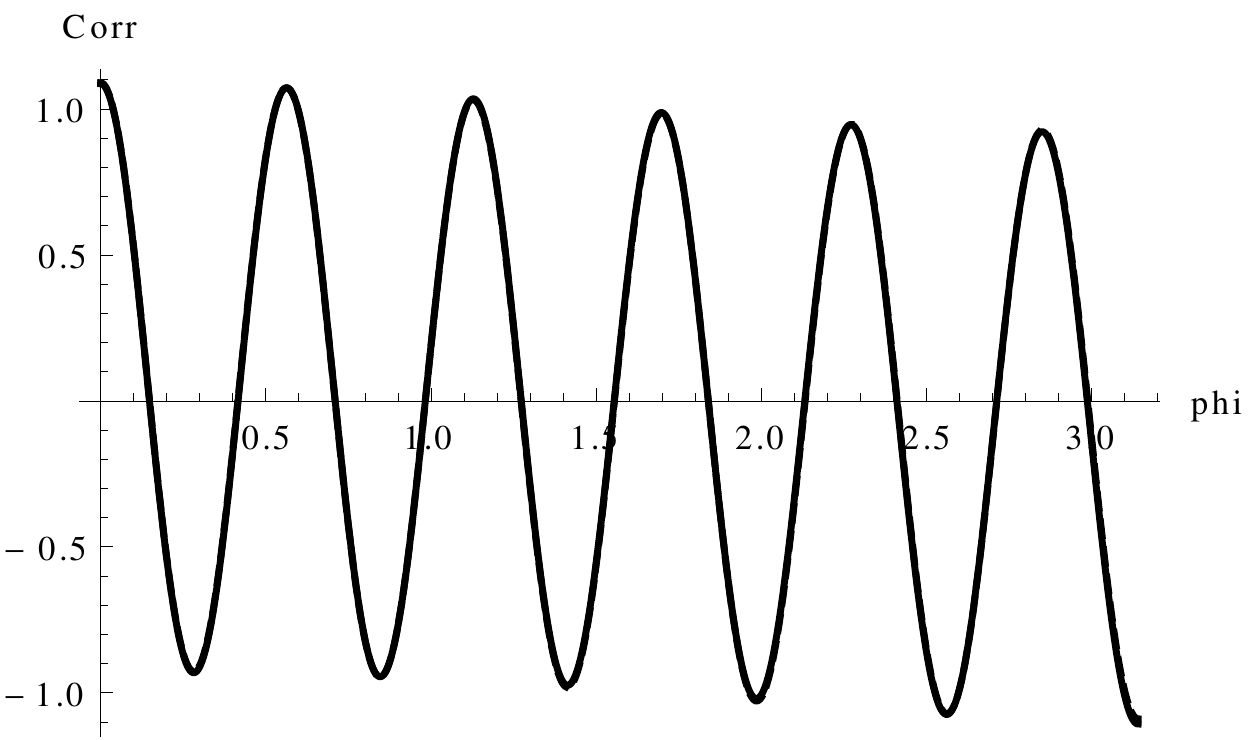}}
\caption{\label{N11} solid line: exact result; dashed line: large $N$}
\end{center}
\end{figure}

\begin{figure}[h!]
\begin{center}
\subfigure[N=29, t=0.3]{\includegraphics[scale=0.4]{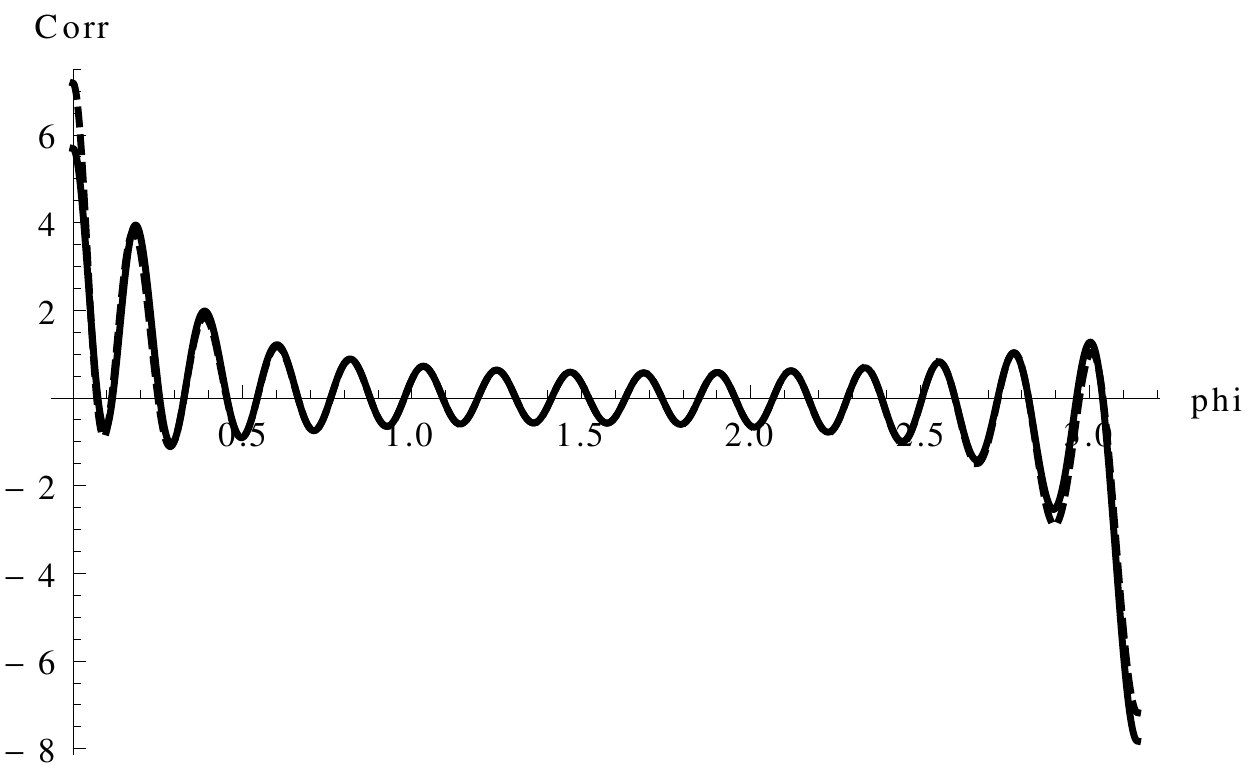}}
\subfigure[N=29, t=1]{\includegraphics[scale=0.4]{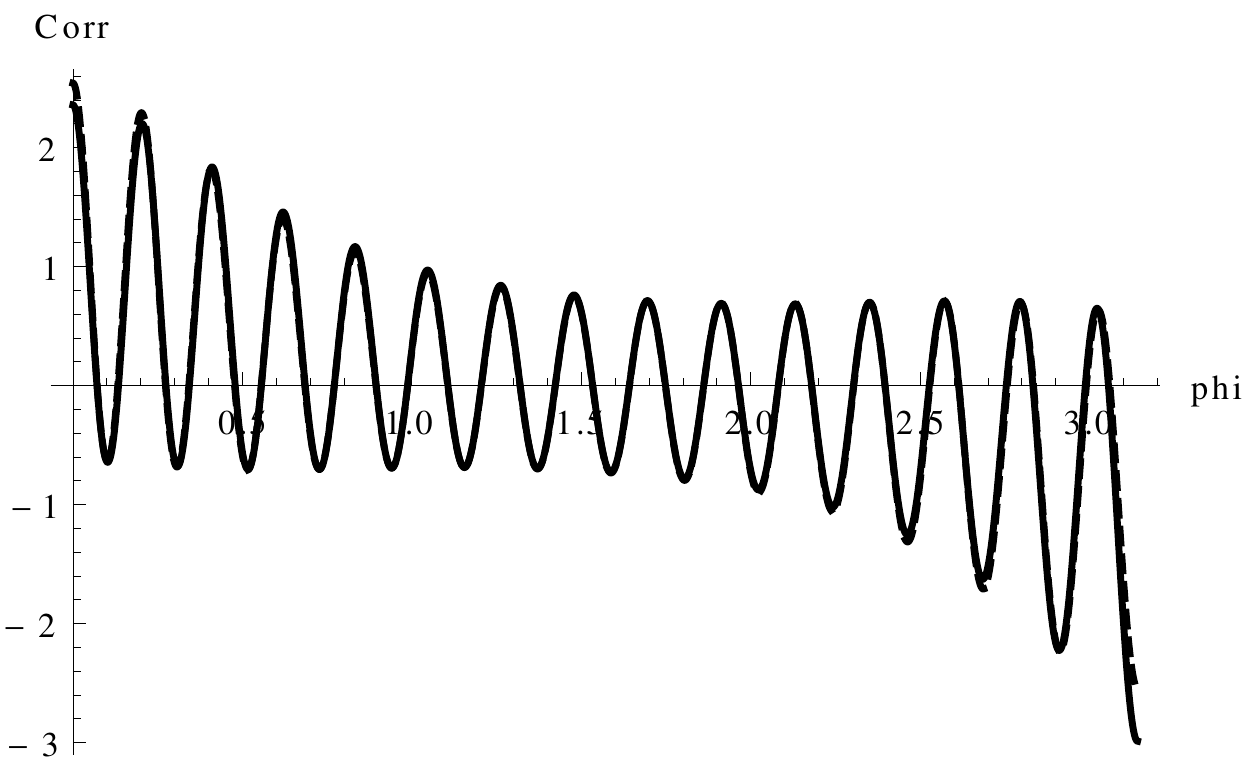}}
\subfigure[N=29, t=5]{\includegraphics[scale=0.4]{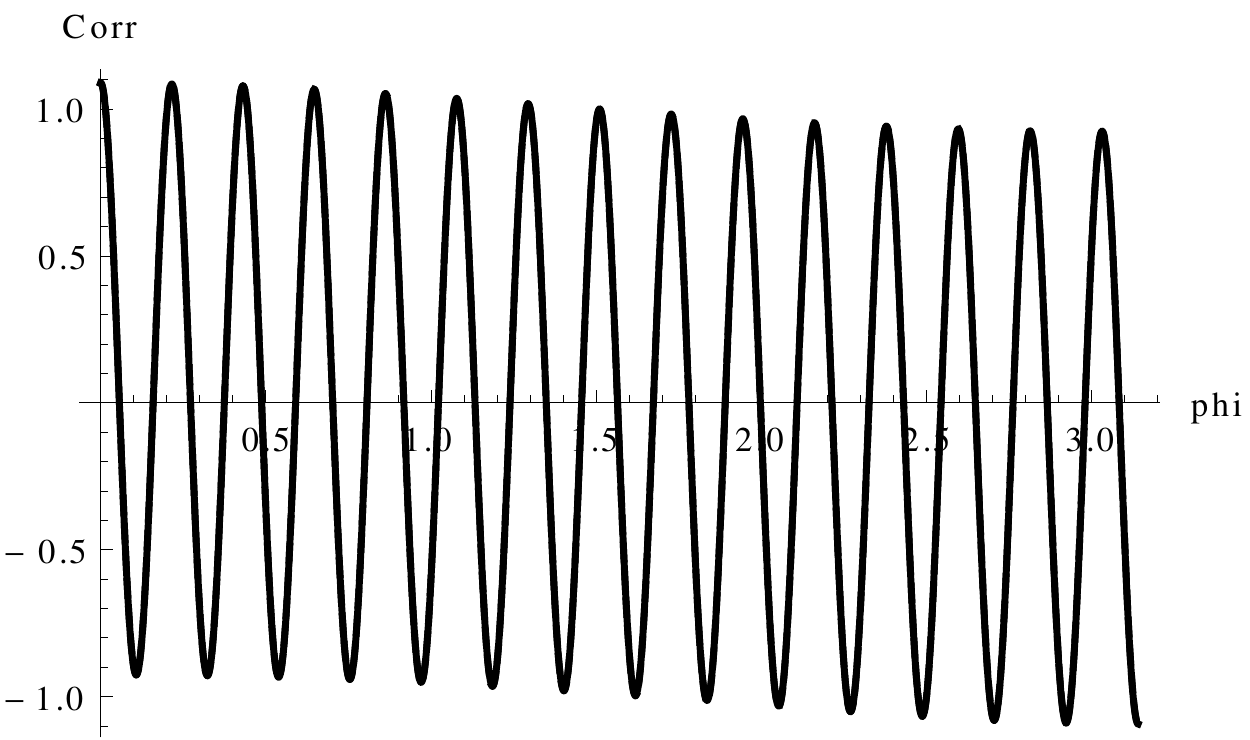}}
\caption{\label{N29} solid line: exact result; dashed line: large $N$}
\end{center}
\end{figure}

\subsubsection{Some 4D examples}

I now turn to 4D and report on some numerical simulation done in order to see
qualitatively whether, overall, the data looks similar to the 2D case. I 
only wish to confirm that also in the 4D case there are no signs of a large
$N$ transition. I do not aim here for anything
quantitative and am content with low numerical precision.  

The connected single-eigenvalue distribution for two Polyakov loops in 
4D will be a function of $\alpha$ and $\beta$ similarly to the 2D case.
For finite $N$, there is no reason for this function to depend only on the
angle difference. The $Z(N)$ symmetry only provides invariance under simultaneous
shifts of $\alpha$ and $\beta$ by $2\pi k/N$. Initial simulations were 
done collecting two dimensional histograms in the $\alpha,\beta$ plane.
Is was found that within practical numerical accuracy collapsing the
histograms along constant $\alpha-\beta$ lines did not loose any information.
This means that we may as well consider the following finite-$N$ definition of
$\rho^{(2)}$:
\begin{equation}
\rho^{(2)}(\alpha-\beta)=
\frac{N}{2\pi}\int_{-\pi/N}^{\pi/N} d\theta 
\langle \rho^{(1)}_1(\alpha+\theta)\rho^{(1)}_2(\beta+\theta)\rangle_c 
\end{equation}
$\rho^{(2)}$ depends only on the angle difference on account of the $Z(N)$ symmetry.

In order to eliminate the UV divergences in 4D, the gauge field configurations
were smeared. Unlike in previous work~\cite{smearing}, 
gauge fields along the direction separating
the Polyakov loops were left unsmeared; smearing was only done for gauge fields 
tangent to all orthogonal three-spaces. Thus, the gauge fields 
entering the Polyakov loop are smeared. This is enough to 
remove the perimeter divergence,
as can be seen from the formula of a massless propagator smeared in the above manner 
at infinite volume:
\begin{equation}
G(x,s)=\int \frac{d^4 p}{(2\pi)^4} e^{ip_4 x_4 +i{\vec p}\cdot {\vec x}}
\frac{e^{-2s {\vec p}^2}}{p_4^2+{\vec p}^2}
\end{equation}
One has then $G(0,s)=\frac{1}{16 \pi s}$ and the short 
distance singularity is regulated away.
This removes the perimeter divergence at the cost of a 
dependence on the new scale $\sqrt{s}$. 
At the level of Feynman diagrams it is obvious that this 
eliminates all UV perimeter divergences to any finite order, because the extra
diagrams smearing introduces have a tree structure, reflecting the determinism of the 
smearing equation, which, in continuum notation reads:
\begin{equation}
F_{is}=D^{\rm adjoint}_j F_{ij}
\end{equation}
Like in~\cite{smearing}, $s$ is a coordinate along a new direction.
$i,j$ label directions orthogonal to the
direction of separation between the two Polyakov loops.
$D$ and $F$ are the covariant derivative
and field strength respectively. 
The three dimensional character of the smearing means that the quantities 
${\cal F}_R(l)$ are $s$-independent on account of the limit $r\to\infty$ which projects
on the ground state of the relevant Hamiltonian. Smearing only
affects the (regularized) matrix element of the operator between the 
singlet $Z(N)$ ground state and the nontrivial $Z(N)$ ground state.

In simulations we employed $s=0.25$ in lattice units, which is a 
moderate amount of smearing, 
found adequate in the study of contractible Wilson loops~\cite{rect_w_loops}. 

The figures below show results from a lattice volume of $12^4$ with $N=29$ at 
inverse 't Hooft couplings, $b=0.360,0.365,0.370$ and separation $r=1,2,3$ in
lattice units. We see that at fixed $b$ the general behavior resembles the
analytical results in 2D with $t$ increasing with $r$. One also sees a trend
of increase in the difference from 2D as the angle difference increases.
Only the angle difference range of $(0,\pi)$ is plotted on account of the
symmetry under simultaneous sign switch of $\alpha,\beta$. 
As $b$ increases the effective
$t$ decreases, as expected on account of asymptotic freedom. 
The errors on the MC data are of the order of 10 percent but 
cannot be reliably estimated. 

Each figure shows, in addition to raw data, a smoothed curve obtained by a cubic spline
smoothing method. The method of smoothing consists of a minimization
of a weighted combination of some average of curve curvature and deviation from the data. 
The smoothing procedure is quite ad-hoc, and only serves to produce curves to guide the eye.

\begin{figure}[h!]
\begin{center}
\subfigure[N=29, r=1, b=0.36]{\includegraphics[scale=0.4]{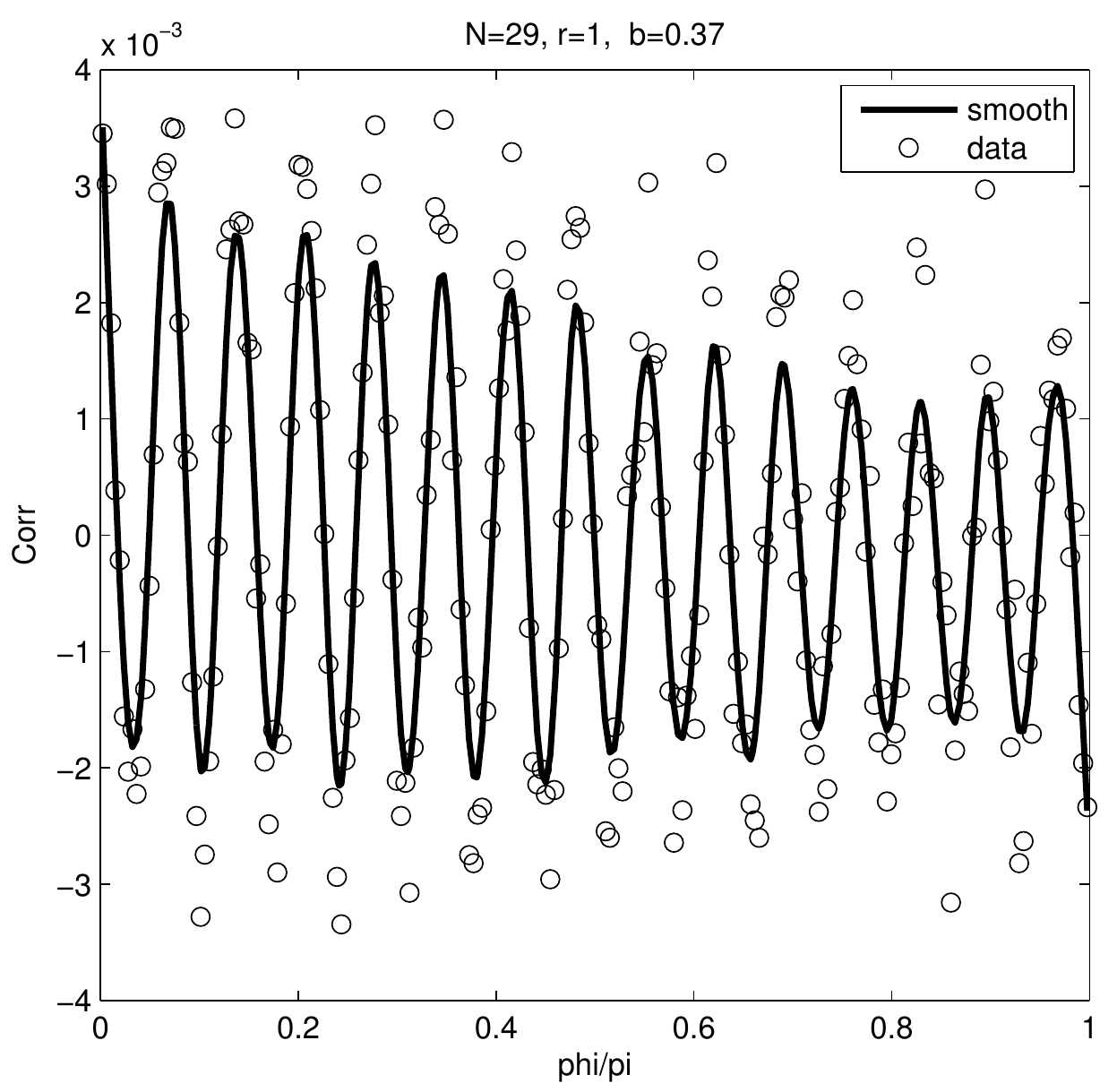}}
\subfigure[N=29, r=2, b=0.36]{\includegraphics[scale=0.4]{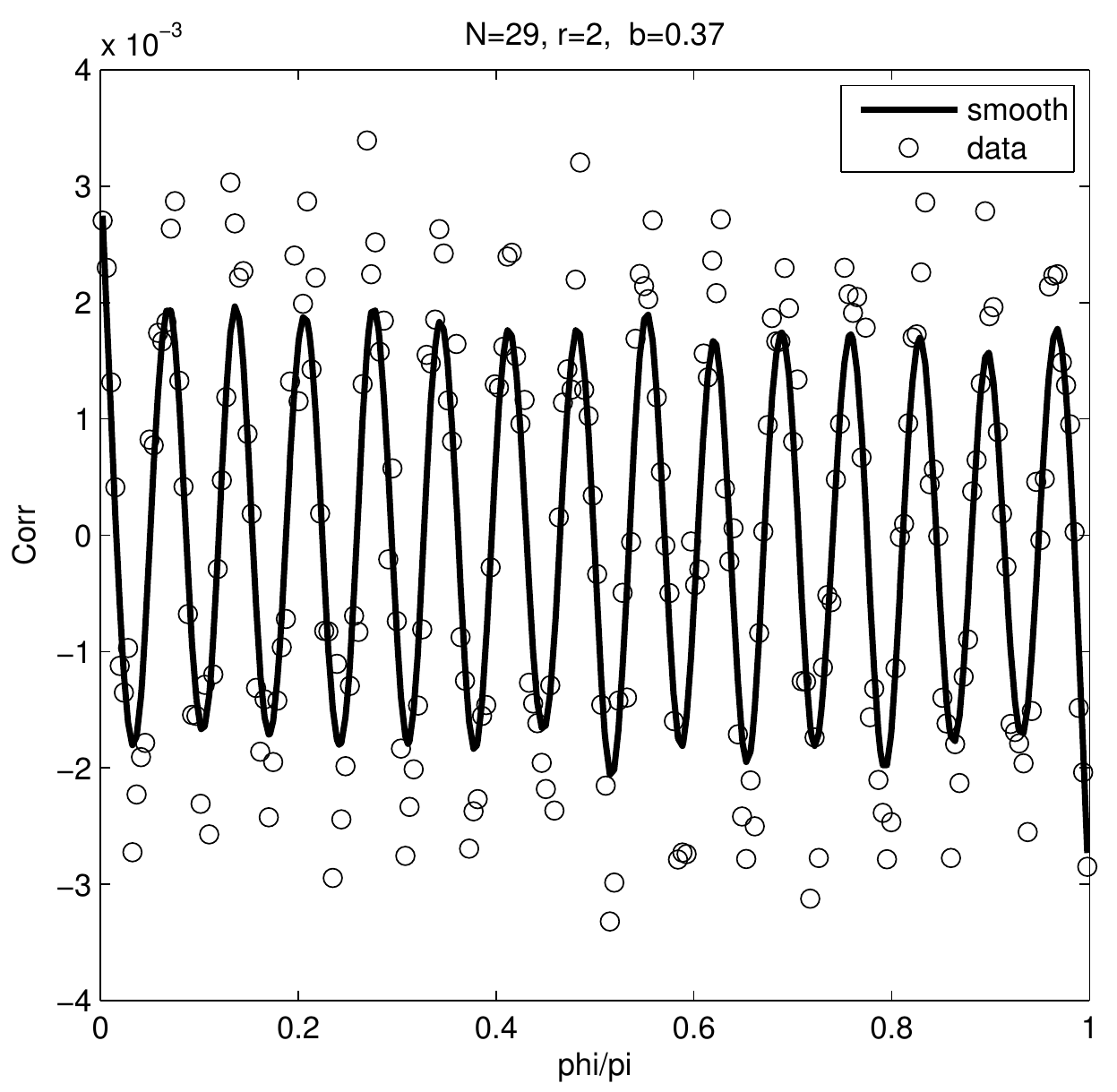}}
\subfigure[N=29, r=3, b=0.36]{\includegraphics[scale=0.4]{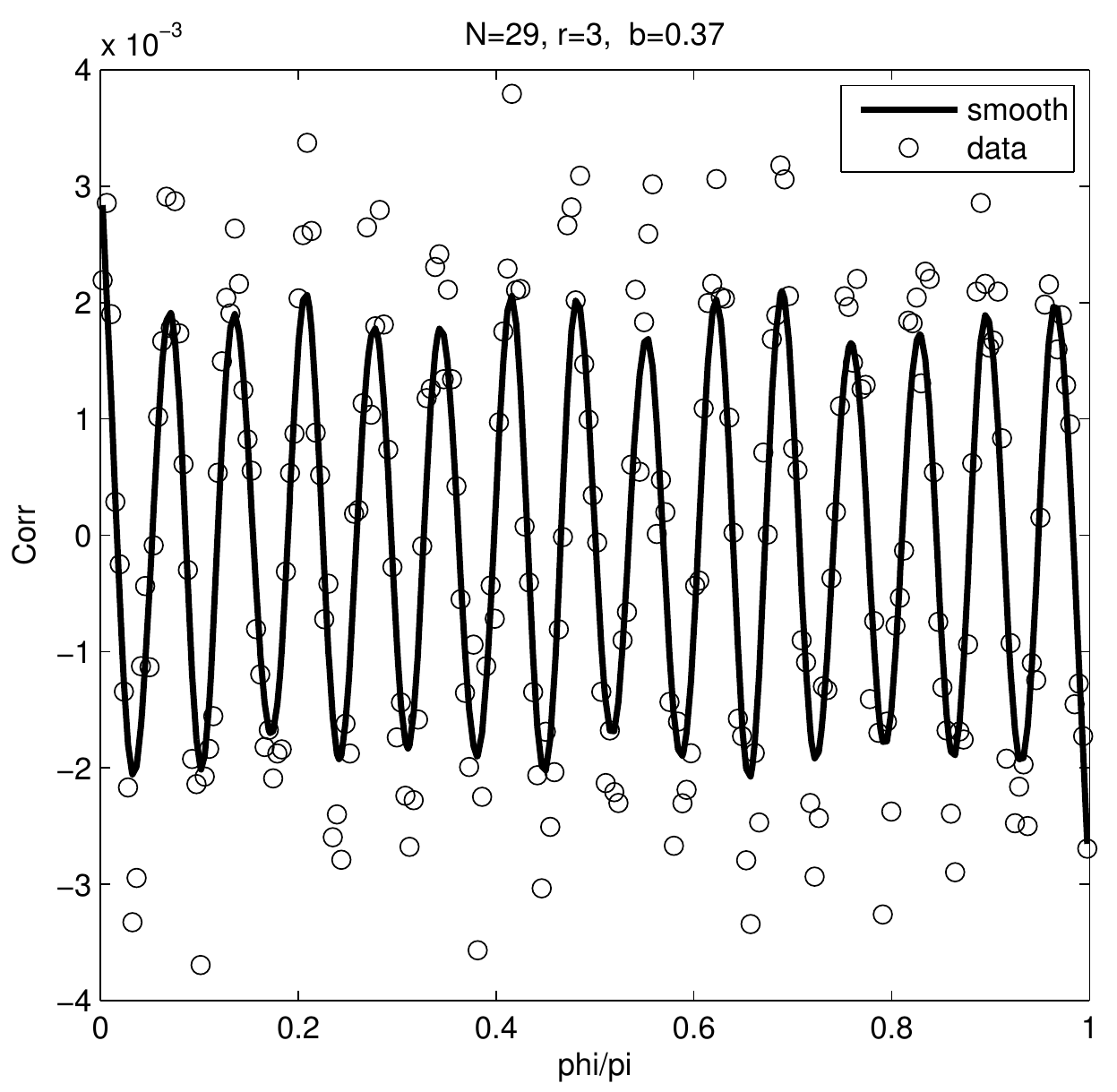}}
\caption{\label{N29mcb36} MC data on $12^4$ at $b=0.36$ and at smearing $s=0.25$}
\end{center}
\end{figure}

\begin{figure}[h!]
\begin{center}
\subfigure[N=29, r=1, b=0.365]{\includegraphics[scale=0.4]{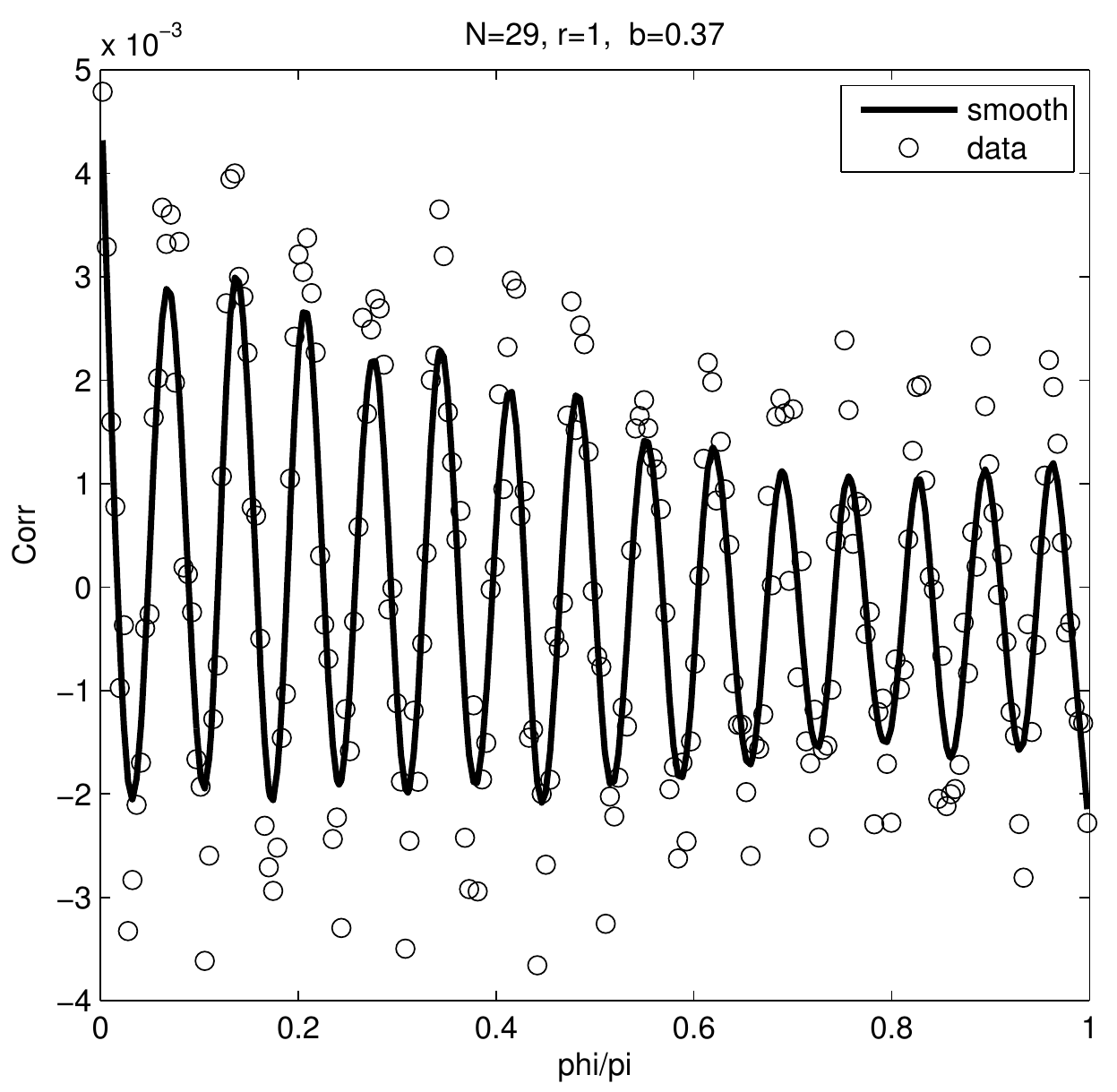}}
\subfigure[N=29, r=2, b=0.365]{\includegraphics[scale=0.4]{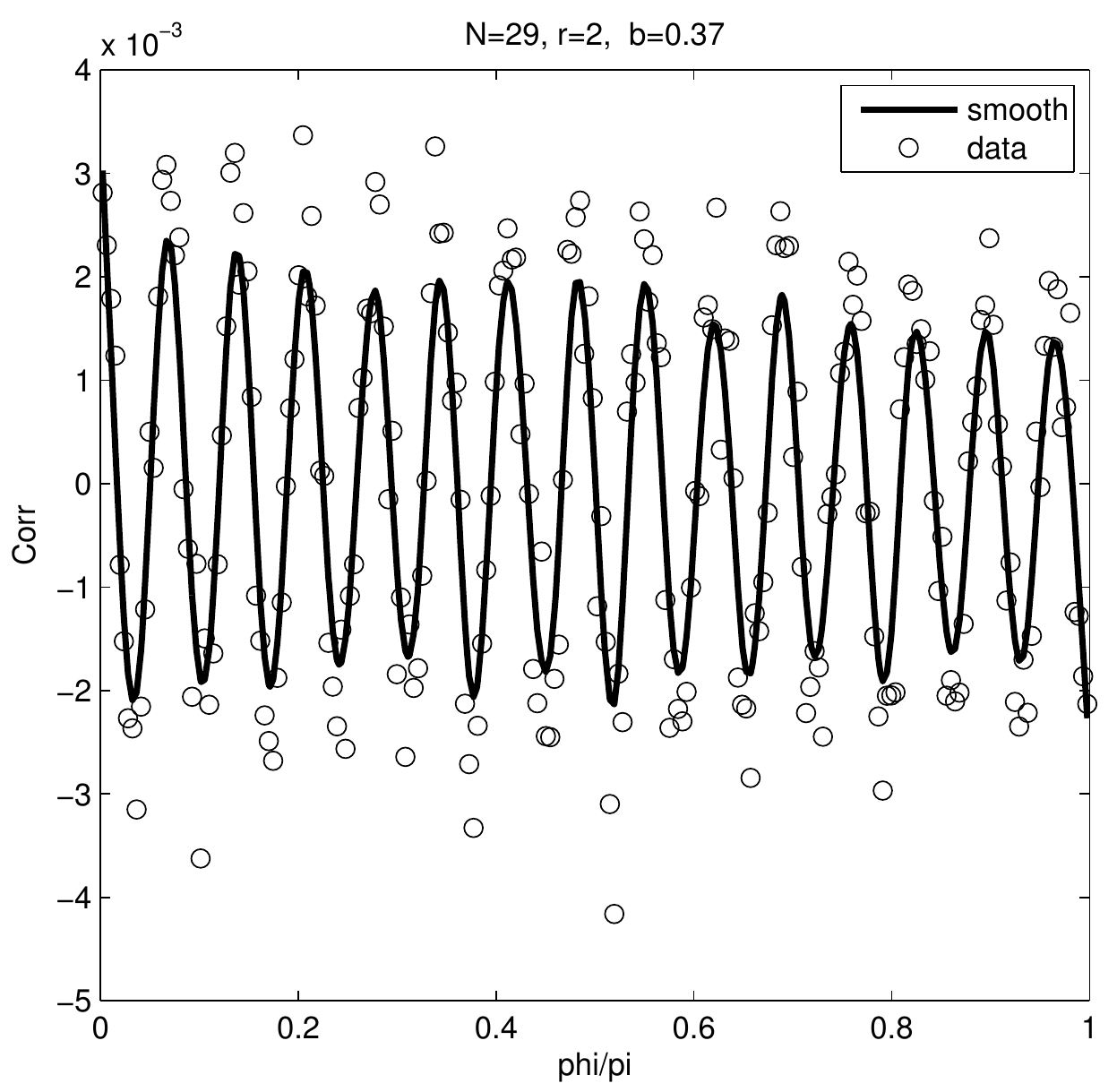}}
\subfigure[N=29, r=3, b=0.365]{\includegraphics[scale=0.4]{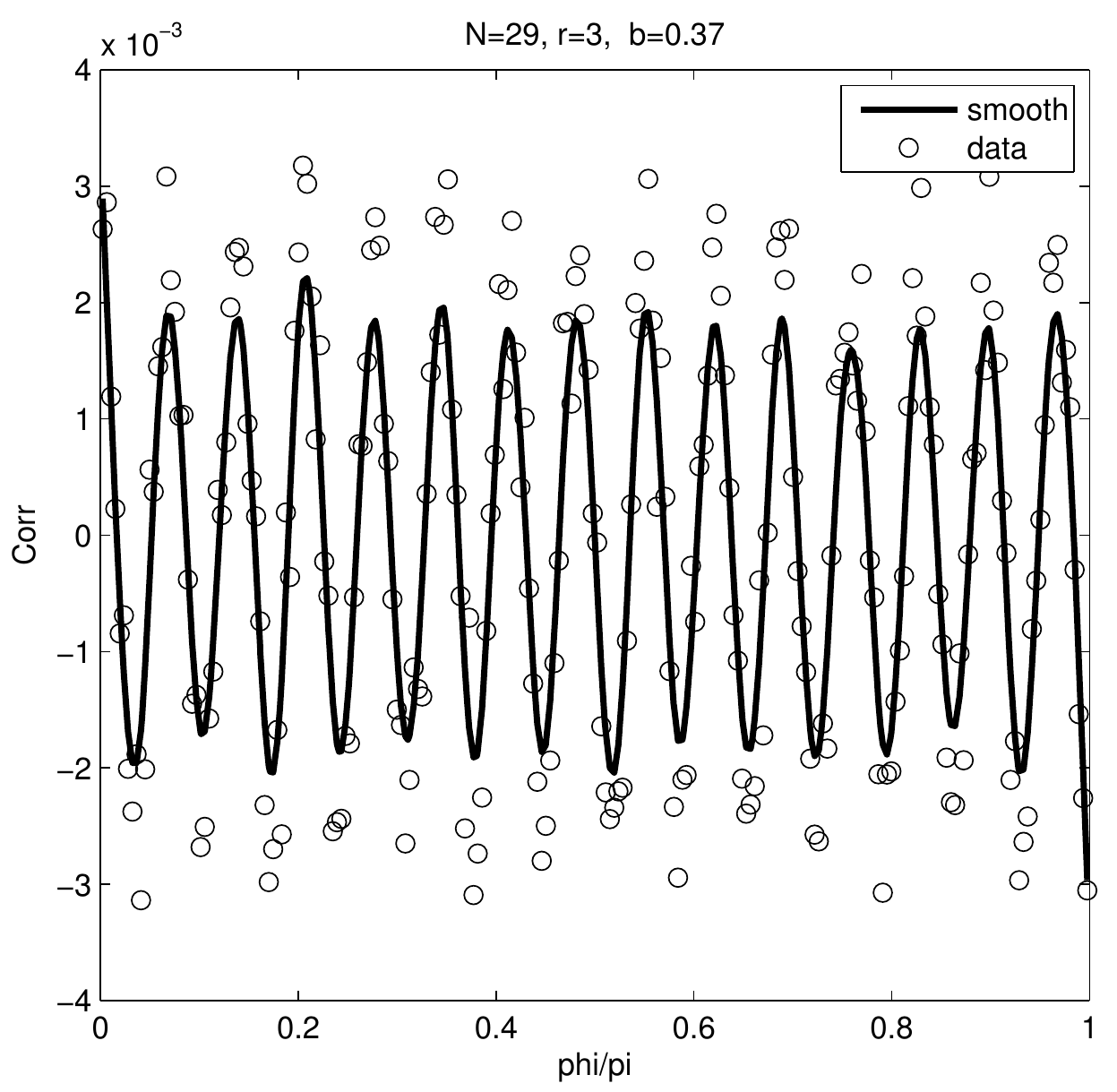}}
\caption{\label{N29mcb365} MC data on $12^4$ at $b=0.365$ and at smearing $s=0.25$}
\end{center}
\end{figure}

\begin{figure}[h!]
\begin{center}
\subfigure[N=29, r=1, b=0.37]{\includegraphics[scale=0.4]{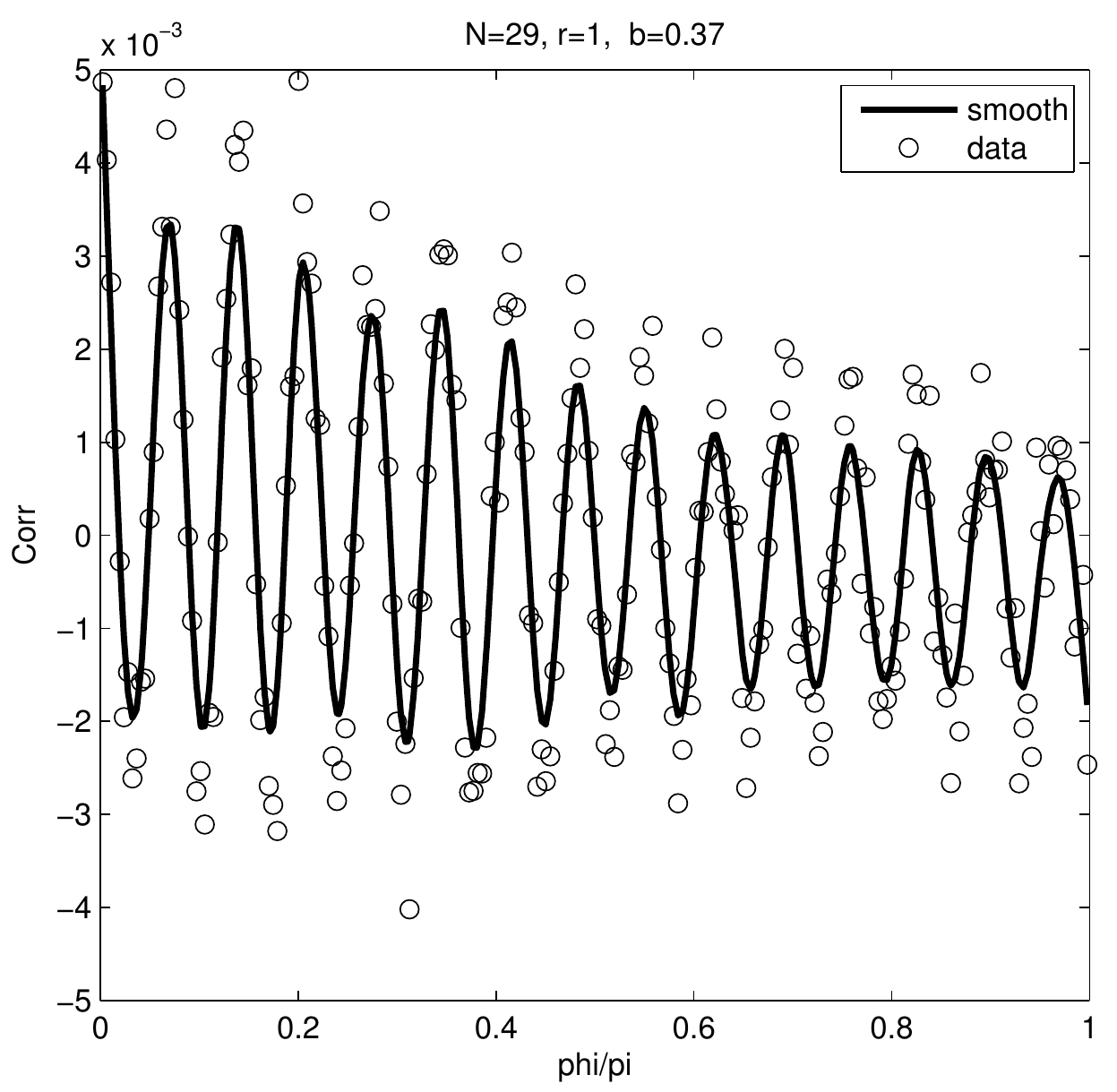}}
\subfigure[N=29, r=2, b=0.37]{\includegraphics[scale=0.4]{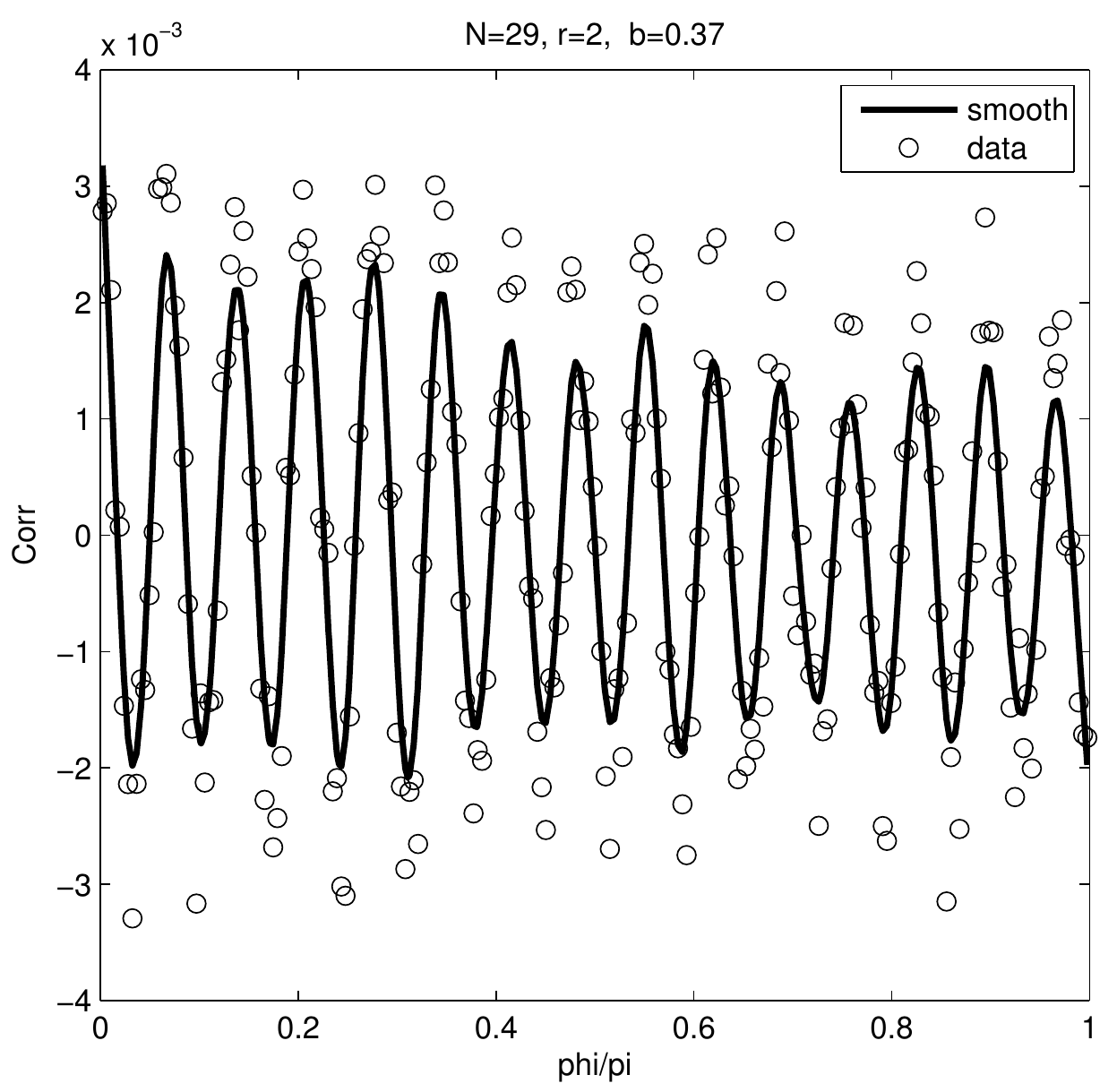}}
\subfigure[N=29, r=3, b=0.37]{\includegraphics[scale=0.4]{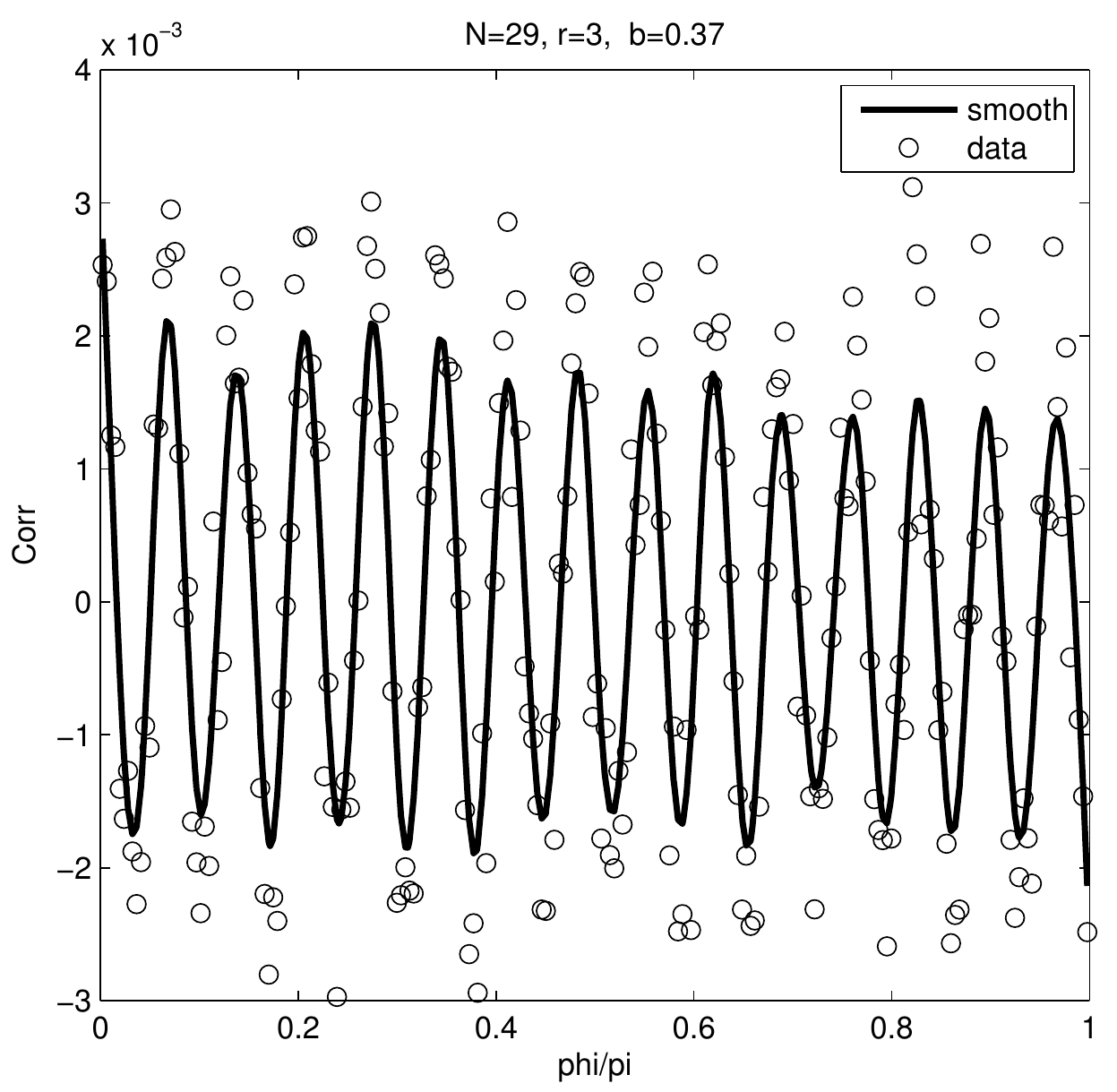}}
\caption{\label{N29mcb37} MC data on $12^4$ at s$b=0.37$ and at smearing $s=0.25$}
\end{center}
\end{figure}

\section{Other possibilities\label{other}}

\subsection{Where to look for a large $N$ phase transition in 4D\label{other_a}}

I discovered no analogues to the large $N$ phase transition found for contractible
Wilson loops in $R^4$ in this case. The reason must be that one cannot make $l$ small. If
we try, we hit a discontinuity already at finite $N$; this
discontinuity becomes quite spectacular at $N=\infty$~\cite{thorn}.
There seems to be no way to qualitatively distinguish between the finite $N$
discontinuity and the $N=\infty$ one.

So, we would like to work in a metastable phase where $l < \frac{1}{T_c}$ but $Z(N)$
is still a good symmetry. More mathematically, we want 
to analytically continue in $l$ from
the low temperature to the high temperature phase. As pointed out in 
~\cite{polchinski} there
is a way to define also a string description there. The instability in the string theory
occurs when some winding string states go tachyonic~\cite{atick_witten}.\footnote{
A simple way to understand winding states is by bosonizing the two dimensional
scalar field describing the compactified dimension on a cylinder~\cite{neuberger-msc}.
The fermions one gets are solitons of the original theory and their integer 
charges under $U(1)$ and $U(1)_5$ are given by the right and left winding numbers.
The constraint on the closed string states that ties the left and right movers ends up
leaving one extra integer labeling the string modes. For a small circumference, 
nontrivial winding modes have positive mass squares, overcoming the negative additive 
contribution reflecting the ordinary tachyon, but for a larger circumference some of the
winding modes have negative mass square. The smallest non-zero windings cross
the tachyonic threshold at the Hagedorn temperature.}

Numerically, one might try to go into the metastable phase by using quenching. 
Originally quenching was introduced as a device to maintain the global $Z^4(N)$
of $T^4$ and reduction to zero volume~\cite{qek}, 
but the idea was flawed~\cite{bringoltz}. 
The flaw was
that it still left alive an annealed mechanism for breaking 
the $Z^4(N)$ to some proper subgroup.
Without respecting full $Z^4(N)$ symmetry, reduction fails. 
However, when
a single direction is compactified, the preservation of the single $Z(N)$ would
not suffer from this flaw. To be sure, we choose prime $N$ 
because then, unlike $Z^4(N)$, $Z(N)$ has not proper subgroups. 

The point would be to check numerically whether the condensation of winding states, 
which would occur beyond the Hagedorn temperature, is something that occurs in 
the analytically continued field theory at finite $N$ as a phase transition (continuous
or not). The alternative might be that there is no such ordinary phase transition,
since $l$ is small enough to enter a field theoretic perturbative regime where
string theory of any traditional sort is inapplicable. Then, on the basis
of analogy with the contractible Wilson loops, one would guess that a large $N$
phase transition would develop in the two point single eigenvalue correlation function.
The investigation of this is left for the future.

The large $N$ phase transition for contractible loops is seen only when considering
simultaneously many irreducible representations. They may be viewed as coming from
multiple windings of the boundary of the loop. Since the loop is contractible, winding numbers are not 
conserved. In the Polyakov case they are, at least for windings
between 0 and $N-1$. 

The analytic continuation from $lT_c > 1$ to $lT_c <1$ should
provide a way to compute (for small $r$ and $l$) $W(l,r)$ from the two Polyakov loop 
correlator directly and for arbitrary irreducible representations. The analytic continuation
would amount to expanding around the one-loop unstable saddle point, given by
\begin{equation}
P(x) =\frac{1}{d_R}\chi_R[{\rm diag}(e^{i\frac{2\pi j}{N}})]
\end{equation}
For odd $N$, $j=0,..,N-1$, where ${\rm diag}$ indicates a diagonal matrix with
the listed elements on its diagonal.  This configuration is $Z(N)$ invariant, but
unstable at one loop order.

\subsection{Correlations of three Polyakov loops\label{other_b}}

There are two ingredients in 2D YM: one is the ``propagator'' defining the
cylinder with fixed circular boundaries and the other is the ``vertex'' 
which sews together three boundaries~\cite{gross_matytsin}. 
This indicates that it would be
of interest to study the connected correlation function of three Polyakov 
loops. 

The simplest example is to take 2 Polyakov loops in the
fundamental and a third in the irreducible representation
made by combining two anti-fundamentals into a symmetric or anti-symmetric
irreducible representation. Take $N\ge 5$ and odd. The three
loops are positioned at distinct locations in $R^3$. 
In Euclidean space, using a different
slicing this looks like a finite temperature setting for one 
among the many possible generalizations to 
large $N$ of $N=3$ baryons. Three infinitely heavy quarks are
connected by a V-shaped string configuration.

One could go to 
Minkowski space and endow these locations with zero masses, 
forcing them to evolve in time at the velocity of light. 
For open strings in Minkowski space such a
situation was looked at in the context of cosmic strings. 
For $N=3$ this was considered in several papers, but
the string tensions were taken to be equal~\cite{kibble}. 
An effective string 
theory valid at large separations would need to handle
a case where one couples two strings of the same tension to one
of a different tension. 
It seems to me that string
tension considerations would favor a V-shaped arrangement of ``fundamental''
strings. It would be interesting to apply the methods of
effective string theory to this setup. Assuming the V-shape, in the field theory
there would be a coupling associated with the vertex of the V. 
For large $N$ it would go as $\frac{g}{N^4}$ with a finite $g$. 

Back to fixed sources, the three point function of Polyakov loops for the 
antisymmetric case is given by:
\begin{equation}
\frac{2}{N^3(N-1)}
\langle Tr(U_{P_1})  [ Tr(U_{P_2}^{\dagger 2}) - (Tr (U^\dagger_{P_2}))^2 ]Tr (U_{P_3})\rangle (r_a,r_b,\theta)
\end{equation}
Here, $P_1$ is at $(0,0,0)$, $P_2$ is at $(r_a,0,0)$ and $P_3$ 
is at $(r_a+r_b \cos\theta, r_b \sin\theta, 0)$ with $r_a\ne r_b$ and $r_a,r_b > 0$.
The new ingredient is the presence of corners. In the case of rectangular
Wilson loops, corners may change the rules of effective string theory, by
exhibiting a field theoretical dependence on loop sides which is not
exponentially suppressed even for
asymptotically large loops. Here, the same question 
can be addressed in a different
set-up. On a hypercubic lattice only $\theta$ 
values which are multiples of $\frac{\pi}{2}$ 
are accessible.
Numerically there would be high noise problems, but it is worth a try.
One could then get back at our main theme, and consider the
connected three point function of the eigenvalues of the three loops. 
To search for large $N$ phase transitions one would need to look at 
three point connected correlations $\rho^{(3)}$, depending on two eigenvalue-angle 
differences 
at infinite $N$. 

It might be of interest to consider the problem of colliding two same direction
wound Polyakov loops in Minkowski space. The three point vertex 
would enter twice to produce a two to two particle scattering dominated by the
exchange of the symmetric and antisymmetric long strings with masses
above and below threshold. The distribution of the excited string
modes of the two separate outgoing strings might provide a thought experiment 
reminiscent to the
Bjorken model for high energy nucleus-nucleus collisions~\cite{bjorken}.

\section{Summary\label{summary}}

The correlations among single eigenvalue distributions associated
with various Polyakov loops has been studied for the simplest arrangement
and found to provide no large-$N$ generated non-analyticities. 
The results might be of some interest in random matrix theory. 
One needs the interplay between different windings to get large $N$ phase transitions 
and also a perturbative regime. One idea was to somehow analytically continue 
in $l$ to $lT_c \ll 1$ and follow the evolution of the single eigenvalue
distribution of a Polyakov loop as a function of $l$. The other was
to construct arrangements involving mixtures of Polyakov loops of different
winding numbers. 

\begin{acknowledgments}
I acknowledge partial support by the
DOE under grant number DE-FG02-01ER41165. I am grateful for support under the
Weston visiting scientist program at the Weizmann Institute 
in the Faculty of Physics.  I have benefited from communications with 
Zohar Komargodski and Owe Philipsen.
\end{acknowledgments}

\end{document}